\documentclass[aps,prc,preprint,superscriptaddress]{revtex4-2}

\usepackage{graphicx}
\usepackage{dcolumn}
\usepackage{amsmath}
\usepackage{amssymb}
\usepackage{siunitx}
\usepackage{dcolumn}
\usepackage[sort&compress]{natbib}
\usepackage{color}

\begin{document}


\title{Isochronous mass spectrometry at the RIKEN Rare-RI Ring facility}


\author{D.~Nagae}
\email[]{nagae.d.aa@m.titech.ac.jp}
\affiliation{RIKEN Nishina Center, RIKEN, 2-1 Hirosawa, Wako-shi, Saitama 351-0198, Japan}
\affiliation{Research Center for SuperHeavy Elements, Kyushu University, 744 Motooka, Nishi-ku, Fukuoka-shi, Fukuoka 819-0395, Japan}
\affiliation{Laboratory for Zero-Carbon Energy, Tokyo Institute of Technology, 2-12-1 Ookayama, Meguro-ku, Tokyo 152-8550, Japan}
\author{S.~Omika}
\affiliation{Department of Physics, Saitama University, Shimo-Okubo 255, Sakura-ku Saitama-shi, Saitama, 338-8570 Japan}
\affiliation{RIKEN Nishina Center, RIKEN, 2-1 Hirosawa, Wako-shi, Saitama 351-0198, Japan}
\author{Y.~Abe}
\affiliation{RIKEN Nishina Center, RIKEN, 2-1 Hirosawa, Wako-shi, Saitama 351-0198, Japan}
\author{Y.~Yamaguchi}
\affiliation{RIKEN Nishina Center, RIKEN, 2-1 Hirosawa, Wako-shi, Saitama 351-0198, Japan}
\author{F.~Suzaki}
\affiliation{RIKEN Nishina Center, RIKEN, 2-1 Hirosawa, Wako-shi, Saitama 351-0198, Japan}
\affiliation{Advanced Science Research Center, Japan Atomic Energy Agency, 2-4 Shirakata, Tokai-mura, Naka-gun, Ibaraki 319-1195, Japan}
\author{K.~Wakayama}
\affiliation{Department of Physics, Saitama University, Shimo-Okubo 255, Sakura-ku Saitama-shi, Saitama, 338-8570 Japan}
\author{N.~Tadano}
\affiliation{Department of Physics, Saitama University, Shimo-Okubo 255, Sakura-ku Saitama-shi, Saitama, 338-8570 Japan}
\author{R.~Igosawa}
\affiliation{Department of Physics, Saitama University, Shimo-Okubo 255, Sakura-ku Saitama-shi, Saitama, 338-8570 Japan}
\author{K.~Inomata}
\affiliation{Department of Physics, Saitama University, Shimo-Okubo 255, Sakura-ku Saitama-shi, Saitama, 338-8570 Japan}
\author{H.~Arakawa}
\affiliation{Department of Physics, Saitama University, Shimo-Okubo 255, Sakura-ku Saitama-shi, Saitama, 338-8570 Japan}
\author{K.~Nishimuro}
\affiliation{Department of Physics, Saitama University, Shimo-Okubo 255, Sakura-ku Saitama-shi, Saitama, 338-8570 Japan}
\author{T.~Fujii}
\affiliation{Department of Physics, Saitama University, Shimo-Okubo 255, Sakura-ku Saitama-shi, Saitama, 338-8570 Japan}
\author{T.~Mitsui}
\affiliation{Department of Physics, Saitama University, Shimo-Okubo 255, Sakura-ku Saitama-shi, Saitama, 338-8570 Japan}
\author{T.~Yamaguchi}
\affiliation{Department of Physics, Saitama University, Shimo-Okubo 255, Sakura-ku Saitama-shi, Saitama, 338-8570 Japan}
\affiliation{Tomonaga Center for the History of the Universe, University of Tsukuba, 1-1-1 Tennodai, Tsukuba-shi, Ibaraki, 305-8571, Japan}
\author{T.~Suzuki}
\affiliation{Department of Physics, Saitama University, Shimo-Okubo 255, Sakura-ku Saitama-shi, Saitama, 338-8570 Japan}
\author{S.~Suzuki}
\affiliation{Institute of Physics, University of Tsukuba, 1-1-1 Tennodai, Tsukuba-shi, Ibaraki 305-8571, Japan}
\author{T.~Moriguchi}
\affiliation{Institute of Physics, University of Tsukuba, 1-1-1 Tennodai, Tsukuba-shi, Ibaraki 305-8571, Japan}
\author{M.~Amano}
\affiliation{Institute of Physics, University of Tsukuba, 1-1-1 Tennodai, Tsukuba-shi, Ibaraki 305-8571, Japan}
\author{D.~Kamioka}
\affiliation{Institute of Physics, University of Tsukuba, 1-1-1 Tennodai, Tsukuba-shi, Ibaraki 305-8571, Japan}
\author{A.~Ozawa}
\affiliation{Institute of Physics, University of Tsukuba, 1-1-1 Tennodai, Tsukuba-shi, Ibaraki 305-8571, Japan}
\author{S.~Naimi}
\affiliation{RIKEN Nishina Center, RIKEN, 2-1 Hirosawa, Wako-shi, Saitama 351-0198, Japan}
\author{Z.~Ge}
\affiliation{RIKEN Nishina Center, RIKEN, 2-1 Hirosawa, Wako-shi, Saitama 351-0198, Japan}
\affiliation{Department of Physics, Saitama University, Shimo-Okubo 255, Sakura-ku Saitama-shi, Saitama, 338-8570 Japan}
\author{Y.~Yanagisawa}
\affiliation{RIKEN Nishina Center, RIKEN, 2-1 Hirosawa, Wako-shi, Saitama 351-0198, Japan}
\author{H.~Baba}
\affiliation{RIKEN Nishina Center, RIKEN, 2-1 Hirosawa, Wako-shi, Saitama 351-0198, Japan}
\author{S.~Michimasa}
\affiliation{Center for Nuclear Study, The University of Tokyo, 2-1 Hirosawa, Wako, Saitama 351-0198, Japan}
\author{S.~Ota}
\affiliation{Center for Nuclear Study, The University of Tokyo, 2-1 Hirosawa, Wako, Saitama 351-0198, Japan}
\author{G.~Lorusso}
\affiliation{Department of Physics, University of Surrey, Guildford, Surrey GU2 7XH, United Kingdom}
\author{Yu.A.~Litvinov}
\affiliation{GSI Helmholtzzentrum f\"{u}r Schwerionenforschung GmbH, Planckstra{\ss}e 1, 64291 Darmstadt, Germany}
\author{M.~Wakasugi}
\affiliation{RIKEN Nishina Center, RIKEN, 2-1 Hirosawa, Wako-shi, Saitama 351-0198, Japan}
\author{T.~Uesaka}
\affiliation{RIKEN Nishina Center, RIKEN, 2-1 Hirosawa, Wako-shi, Saitama 351-0198, Japan}
\author{Y.~Yano}
\affiliation{RIKEN Nishina Center, RIKEN, 2-1 Hirosawa, Wako-shi, Saitama 351-0198, Japan}


\date{\today}

\begin{abstract}
A dedicated isochronous storage ring, named the Rare-RI Ring, was constructed at the RI Beam Factory of RIKEN,
aiming at precision mass measurements of nuclei located in uncharted territories of the nuclear chart.
The Rare-RI Ring employs the isochronous mass spectrometry technique with the goal to achieve a relative mass precision of $10^{-6}$ within a measurement time of less than 1 ms. 
The performance of the facility was demonstrated through mass measurements of neutron-rich nuclei with well-known masses.
Velocity or magnetic rigidity is measured for every particle prior to its injection into the ring, wherein its revolution time is accurately determined. The latter quantity is used to determine the mass of the particle, while the former one is needed for non-isochronicity corrections.
Mass precisions on the order of $10^{-5}$ were achieved in the first commissioning, which demonstrates that Rare-RI Ring is a powerful tool for mass spectrometry of short-lived nuclei.
\end{abstract}


\maketitle


\section{Introduction}
Mass is a fundamental quantity of an atomic nucleus which directly reflects its binding energy.
Systematic studies of nuclear masses reveal
 information regarding the evolution of nuclear structure 
such as shell closures, changes of shapes, nucleon-nucleon correlations, weak and electromagnetic interactions, the equation of state for nuclear matter, 
and borders of nuclear existence~\cite{ty-prog}.
Nuclear masses are particularly important for modeling the processes of synthesis of chemical elements in stellar environments~\cite{Cowan-2021}. Data on nuclear masses are nearly complete for most of the nucleosynthesis processes with the exception of the rapid neutron capture process ($r$ process)~\cite{Horowitz_2019}. 
The $r$ process occurs in violent stellar events characterized by a high neutron density.
Presently favored are binary neutron star mergers, which is supported by the identification of strontium in the lightcurve analysis~\cite{Kilonova-Sr} of so far the only observed kilonova AT2017gfo~\cite{Kilonova,Kilomova-r-process}, which is related to the gravitational wave detection GW170817~\cite{GW170817}.
Numerous rapid captures of free neutrons drive matter towards the neutron drip line.
Therefore, the nuclei involved in the $r$ process are extremely neutron rich and as a rule short lived. 
The neutron capture and photodisintegration reaction rates critically depend on the neutron separation energy, which is the difference of masses of the corresponding parent and daughter nuclei and a free neutron.
Sensitivity studies of $r$-process simulations for uncertainties of nuclear masses~\cite{Sun-2008r,Arcones-2011,Jiang-2021} indicate that masses need to be known with accuracy of better than 100~keV/$c^{2}$.

Various approaches developed for precision mass measurements of exotic nuclei~\cite{ty-prog} include Penning traps~\cite{dilling-penning, bollen-penning-ISOLTRAP, dilling-penning-SHIPTRAP, kolhinen-penning-JYFL, ringle-penning-LEBIT}, multi-reflection time-of-flight mass spectrographs~\cite{wolf-mrtof, schury-mrtof, plass-mrtof, jesch-mrtof}, and heavy-ion storage rings~\cite{yuri-esr-sms, Hausmann-2000,Hausmann-2001, tu-csre-ims}.
Altogether, a vast amount of mass data was provided for both neutron-rich and neutron-deficient nuclei~\cite{ty-prog}.
In some cases uncertainties of the mass values can reach sub-keV level.
However, precise mass measurements of exotic nuclei relevant for the $r$ process remain challenging due to their short lifetimes and vanishingly small production rates. 
Only a few successful measurements have been reported so far; see, e.g.,~\cite{Baruah-2008,Sun-2008b,Atanasov-2015b, Knobel-2016b,knobel-esr-ims,Vilen-2018}.

The RIKEN RI Beam Factory (RIBF) is an accelerator complex consisting of several linacs and cyclotrons~\cite{yano}. The RIBF is currently the most powerful radioactive isotope (RI) beam production facility, giving access to many neutron-rich nuclei in the vicinity of the $r$ process.
To profit from this RIBF capability, a special cyclotron-like storage ring, named the Rare-RI Ring, was constructed at RIBF~\cite{ozawa}.
The Rare-RI Ring is designed to conduct mass measurements by utilizing isochronous mass spectrometry.
The envisioned relative mass precision is $\delta m / m \approx 10^{-6}$. Since the measurement time is approximately $1$~ms, the technique is quick enough to cover basically all relevant nuclear species.

Driving accelerators at other existing heavy-ion storage rings---ESR at GSI~\cite{franzke}, and CSRe at IMP~\cite{xia}---are  heavy-ion synchrotrons. 
Synchrotrons and storage rings are synchronized and operate with bunched beams ejected and/or injected within one revolution in both machines~\cite{Steck-2020}. 
Thus, the entire beam is ejected within several hundred nanoseconds. Neither particle identification nor velocity determination are feasible in the beam-line and have thus to be done inside the ring, which is often challenging~\cite{Geissel-2005, Geissel-2006, Meng-2022, Geissel-2023, Zhang-2023, Zhou-2023, Wang-2023a}.

In contrast, the Rare-RI Ring is coupled to a cyclotron.
The quasicontinuous beam structure makes possible, by utilizing commonly used timing, tracking and ionization detectors, the particle identification as well as velocity measurement of every ion prior to its injection into the ring.
However, the very short injection time effectively picks out only a single ion from the beam.
Since various particles are produced in nuclear reactions at random times, the injection must be synchronized only with an ion of interest, which requires its prior identification.
All of these were realized in a special individual-ion injection method, initially proposed in Ref.~\cite{meshkov}.

In the present study, we report the first commissioning mass measurements using nuclei with well-known masses 
as well as the  developed analysis methods.
The first mass measurements at the storage ring in RIKEN were reported by us in Ref.~\cite{Li-RIKEN}.
That measurement became possible after a thorough characterization and development of the corresponding instrumentation. In the preset work we report the details of the employed experimental technique and the initial commissioning results.

\section{Principle of mass measurements}
In the isochronous mass spectrometry, an isochronous condition (isochronism) of the storage-ring optics is tuned using a reference nucleus whose rest mass $m_{0}$ is well known.
As a result, the revolution times of the reference particles $T_{0}$ with the mass-to-charge ratio $m_{0}/q_{0}$ are constant, and are independent of their momenta.
They are given by
\begin{eqnarray}
T_0 = 2\pi \frac{m_0}{q_0}\frac{1}{B_0},
\end{eqnarray}
where $B_0$ is the  magnetic flux density.
For a nucleus of interest with a mass-to-charge ratio $m_{1}/q_{1} = m_{0}/q_{0} \pm \Delta ( m/q )$, the isochronism is not fulfilled~\cite{Zhang-2016}.
The corresponding revolution times $T_{1}$ depend on particle momenta and the distribution of $T_{1}$ rapidly becomes wider with increasing $\Delta (m/q)$.
It is assumed that ions with the same magnetic rigidities $B\rho$ have identical mean orbit lengths in the ring. 
If one of these ions is a reference, the following equations are satisfied:
\begin{equation}
\label{eq:Brho}
B \rho = \frac{m_{0}}{q_{0}} \beta_{0} \gamma_{0} c = \frac{m_{1}}{q_{1}} \beta_{1} \gamma_{1} c,
\end{equation}
\begin{equation}
\label{eq:betaT}
\beta_{0} T_{0} = \beta_{1} T_{1},
\end{equation}
where $\beta_{0, 1}$ and $\gamma_{0, 1}$ are the velocity in units of light speed $c$ and the relativistic Lorentz factor, respectively. The subscripts 0 and 1 label the reference nucleus and the nucleus of interest, respectively.
The mass-to-charge ratio $m_{1}/q_{1}$ can be derived from Eqs.~(\ref{eq:Brho}) and (\ref{eq:betaT}) as follows:
\begin{equation}
\label{eq:m/q-beta}
\begin{aligned}
\frac{m_{1}}{q_{1}} &= \frac{m_{0}}{q_{0}} \frac{T_{1}}{T_{0}} \frac{\gamma_{0}}{\gamma_{1}} \\ 
                &= \frac{m_{0}}{q_{0}} \frac{T_{1}}{T_{0}} \sqrt{\frac{1-\beta_{1}^{2}}{1-\left\{ \left( \frac{T_{1}}{T_{0}} \right) \beta_{1} \right\}^{2}}}\\
                &= \frac{m_{0}}{q_{0}} \frac{T_{1}'}{T_{0}},
\end{aligned}
\end{equation}
where $T_{1}'$ denotes $T_{1}$ corrected by $\beta_{1}$.
This correction is indispensable and directly affects the attainable mass precision.
The relative differential of $m_{1}/q_{1}$ is given as:
\begin{equation}
\label{eq:error-m/q-beta}
\begin{gathered}
\frac{\delta (m_{1}/q_{1})}{m_{1}/q_{1}} = \frac{\delta (m_{0}/q_{0})}{m_{0}/q_{0}}+\gamma_{0}^{2} \frac{\delta (T_{1}/T_{0})}{T_{1}/T_{0}}+k\frac{\delta \beta_{1}}{\beta_{1}}, \\
k = -\frac{\beta_{1}^{2}}{1-\beta_{1}^{2}}+\left(\frac{T_{1}}{T_{0}}\right)^{2}\frac{\beta_{1}^{2}}{1-(T_{1}/T_{0})^{2}\beta_{1}^{2}}.
\end{gathered}
\end{equation}
The mass of the nucleus of interest is determined with a precision of $10^{-6}$ when $T_{0, 1}$ and $\beta_{1}$ are known to a precision of $10^{-6}$ and $10^{-4}$, respectively.
The coefficient $k$ is on the order of $10^{-2}$ for $\Delta (m/q)/(m_{0}/q_{0})$ of $10^{-2}$. 

$m_{1}/q_{1}$ can as well be obtained by using the revolution time $T_{1}$ corrected by $B\rho$:
\begin{equation}
\label{eq:m/q-brho}
\begin{aligned}
\frac{m_{1}}{q_{1}} &= \frac{m_{0}}{q_{0}} \frac{T_{1}}{T_{0}} \frac{\gamma_{0}}{\gamma_{1}} \\
                   &= \frac{m_{0}}{q_{0}} \frac{T_{1}}{T_{0}}
\sqrt{ 1 +\frac{1-\left( T_{0} /T_{1} \right)^2} {\left( \frac{ \left( m_{0} / q_{0} \right) }{B\rho} c \right)^2 }}\\
                   &= \frac{m_{0}}{q_{0}} \frac{T_{1}''}{T_{0}}.
\end{aligned}
\end{equation}
$T_{1}''$ is the magnetic-rigidity correction to $T_{1}$.
In this case, the relative differential of $m_{1}/q_{1}$ is given as:
\begin{equation}
\label{eq:error-m/q-brho}
\begin{aligned}
\frac{\delta (m_{1}/q_{1})}{m_{1}/q_{1}} = \kappa \frac{\delta (m_{0}/q_{0})}{m_{0}/q_{0}} + \lambda \frac{\delta (T_{1}/T_{0})}{T_{1}/T_{0}}+ \mu \frac{\delta B\rho}{B\rho}.
\end{aligned}
\end{equation}
Here, $\kappa$, $\lambda$, and $\mu$ are given as follows:
\begin{equation}
\label{eq:klm}
\begin{aligned}
\kappa &= \frac{\left\{ \left(m_{0}/q_{0}\right) c\right\}^2}
{\left\{ \left(m_{0}/q_{0}\right) c\right\}^2 + \left(B\rho\right)^2\left\{ 1-\left(T_{0}/T_{1}\right)^2\right\} },\\
\lambda &= \frac{\left\{ \left(m_{0}/q_{0}\right) c\right\}^2 +  \left( B\rho \right)^2}
{\left\{ \left(m_{0}/q_{0}\right) c\right\}^2 + \left(B\rho\right)^2\left\{ 1-\left(T_{0}/T_{1}\right)^2\right\} },\\
\mu &= \frac{\left( B \rho \right)^2  \left\{ 1-\left( T_{0}/T_{1} \right)^2  \right\}}
{\left\{ \left(m_{0}/q_{0}\right) c\right\}^2 + \left(B\rho\right)^2\left\{ 1-\left(T_{0}/T_{1}\right)^2\right\} }.
\end{aligned}
\end{equation}
The coefficients $\kappa$, $\lambda$, and $\mu$ are on the order of 1, 1, and $10^{-2}$, respectively.
Thus, the mass of the nucleus of interest is determined with a precision of $10^{-6}$ when $T_{0, 1}$, and $B\rho$ are determined to a precision of $10^{-6}$ and $10^{-4}$, respectively.
In the Rare-RI Ring, $m_{1}/q_{1}$ is determined by measuring the revolution times $T_{0}$ and $T_{1}$ corrected by $\beta_{1}$ or $B\rho$.

\section{The Rare-RI Ring facility}
\begin{figure}[tb]
\centering
\includegraphics[clip, width=17.2cm]{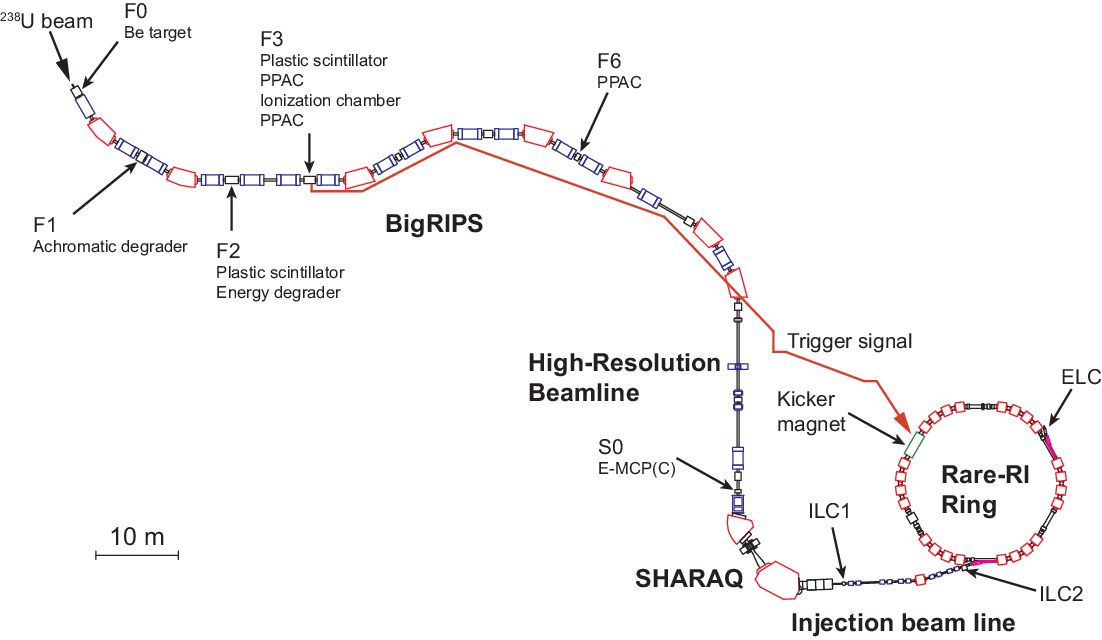}
\caption{Layout of the beamlines and the Rare-RI Ring at RIBF. 
The Rare-RI Ring is located downstream of the SHARAQ spectrometer. 
The red arrow indicates the path of the trigger signals utilizing the high-speed coaxial tube. 
Detectors placed in the long-injection beamline are indicated. 
The E-MCP(C)~\cite{nagae} is a time-of-flight detector equipped with a carbon foil.
}
\label{fig:ribf}
\end{figure}

\begin{figure}[tb]
\centering
\includegraphics[clip, width=12.9cm]{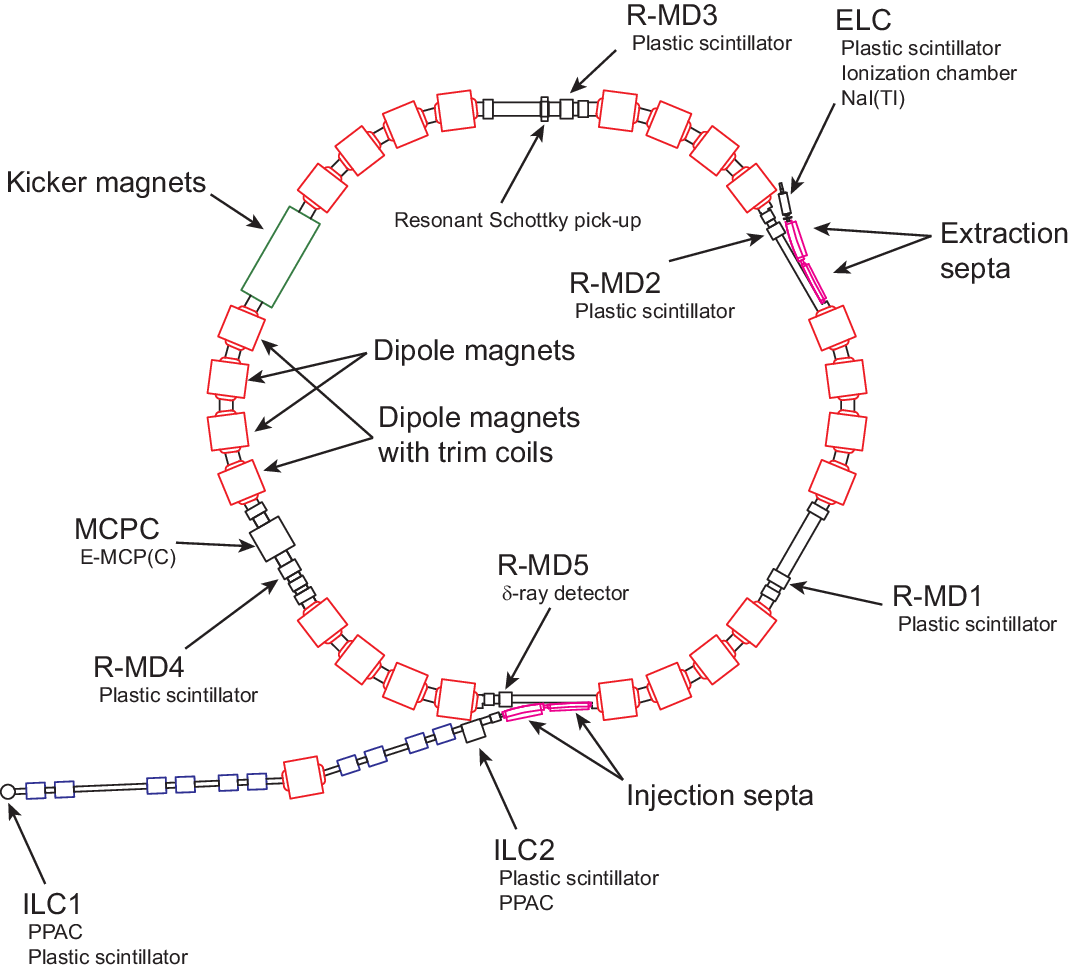}
\caption{Structure of the Rare-RI Ring with the detector setup.
Plastic scintillators placed at R-MD1, R-MD2, and R-MD3 were used to  determine the trajectories of the injected ions.
A plastic scintillator at R-MD4 was used to verify  synchronization of the kicker timing with the particle arrival and the kicking angle.
The E-MCP(C)~\cite{nagae} and $\delta$-ray detector~\cite{omika-delta} were utilized to confirm the storage of particles. 
A resonant Schottky pickup~\cite{suzaki} was installed for precise tuning of the isochronous optics by using long-lived species.
For particle identification, an ionization chamber and a NaI(Tl) counter 
were set up at the ring ejection channel ELC to measure the energy loss and total energy, respectively.}
\label{fig:ring}
\end{figure}

A schematic layout of the Rare-RI Ring and the detector configuration used in the discussed experiment are shown in Figs.~\ref{fig:ribf} and \ref{fig:ring}. 
The Rare-RI Ring is placed downstream of a long-injection beamline starting from the production target at F0 and consisting of the high-acceptance multi-stage fragment separator BigRIPS~\cite{kubo}, the high-resolution beamline~\cite{michimasa}, the SHARAQ spectrometer~\cite{michimasa, uesaka}, and the injection beam line of the Rare-RI Ring itself. 
The long-injection beamline is mandatory to achieve the individual-ion injection~\cite{meshkov} simultaneously with its precise $\beta$ measurement at $10^{-4}$ resolution.

The Rare-RI Ring consists of six sectors, each of which has four dipole magnets. 
To achieve the isochronous ion optical setting, two dipole magnets on both ends of each sector are equipped with ten independent single-turn trim coils.
Isochronism on the order of $10^{-6}$ can be achieved by adjusting the currents in the trim coils.

Because the Rare-RI Ring is coupled to a cyclotron, the injection of a quasicontinuous beam from a cyclotron into the ring is not compatible.
A particle-by-particle injection method into the ring, the so-called individual-ion injection method~\cite{meshkov}, was established.
For its realization, a fast response kicker system~\cite{miura} was developed.
The trigger signal for the kicker system is generated by an ion of interest in the plastic scintillator at the focal plane F3; see Fig.~\ref{fig:ribf}. 
To excite the kicker magnets before the arrival of the ion, the signal is transferred to the kicker system using a high-speed coaxial tube~\cite{omika}.
The achieved signal propagation speed was $98.6$\% of the speed of light. 
Then, a thyratron switch promptly excites the kicker magnets. 
We adopted traveling-wave-type kicker magnets to generate a high magnetic field amplitude with a short rise time.

The momentum acceptance of the Rare-RI Ring is designed to be $\pm 0.5$\%.
The transverse acceptances are $150\pi$ and $30\pi$~mm~mrad in the horizontal and vertical planes, respectively.
The residual gas pressure in the Rare-RI Ring ranges from $2.5 \times 10^{-6}$ to $4.0 \times 10^{-5}$~Pa achieved without baking.
Although the pressure is not at ultrahigh vacuum level, it is sufficient to store particles for approximately $2000$ turns.

\section{First demonstration of mass measurement}
The first mass measurement of exotic nuclei was performed on five fully stripped nuclei, namely $^{79}$As, $^{78}$Ge, $^{77}$Ga, $^{76}$Zn, and $^{75}$Cu, all of which have well-known masses tabulated in the Atomic Mass Evaluation, AME2020~\cite{ame}.
$^{78}$Ge was chosen to be the reference particle for tuning the beam transport and isochronous ion optics of the ring.
The masses, revolution times, and time-of-flights (TOFs) between F3 and S0 of $^{79}$As, $^{78}$Ge, $^{77}$Ga, and $^{76}$Zn were used 
to fix the parameters for the velocity $\beta$ and magnetic rigidity $B\rho$ determination, which is further detailed below.
Subsequently, the mass of $^{75}$Cu was obtained relative to the mass and revolution time of $^{78}$Ge.

The procedure for determining the mass at the Rare-RI Ring is divided into the following five steps:
(1) production of RIs; 
(2) injection of RIs into the ring;
(3) circulation confirmation of RIs in the ring;
(4) fine-tuning of the isochronous optical setting;
(5) storage for several thousands of turns and extraction of RIs. The timing signals at the injection and extraction provide the total TOF.

\subsection{RIs production and events selection}
\label{subsec:PI}
Exotic nuclei near $^{78}$Ge were produced via in-flight fission of $^{238}$U beams impinging at an energy of 345~MeV/nucleon on a $10$-mm-thick $^{9}$Be target.
Fission fragments have a large momentum spread of several percent and only a small fraction of them could be transported and injected into the ring.
By considering an additional momentum-broadening effect in degraders and detectors 
mounted in BigRIPS, a momentum slit at the dispersive focal plane F1 of the BigRIPS was set to $\pm 2$~mm, corresponding to $\pm 0.09$\% in momentum.

Standard particle separation was achieved by a combined analysis of the magnetic rigidity and energy loss in the degraders. 
In this experiment, the magnetic rigidity of the BigRIPS was adjusted to enhance the purity of $^{78}$Ge, which was approximately 40\% at F3.
A wedge-shaped achromatic degrader made of 5-mm-thick aluminum was inserted at F1.
A slit with a $\pm 2$~mm opening at the achromatic focal plane F2 was used to remove undesired contaminants.
An additional energy degrader made of $5.6$-mm-thick aluminum placed at F2 was used for fine tuning the energy of the fragments.
The beam energy of $^{78}$Ge at F3 was approximately 175~MeV/nucleon.

\begin{figure}[tb]
\centering
\includegraphics[clip,width=12.9cm]{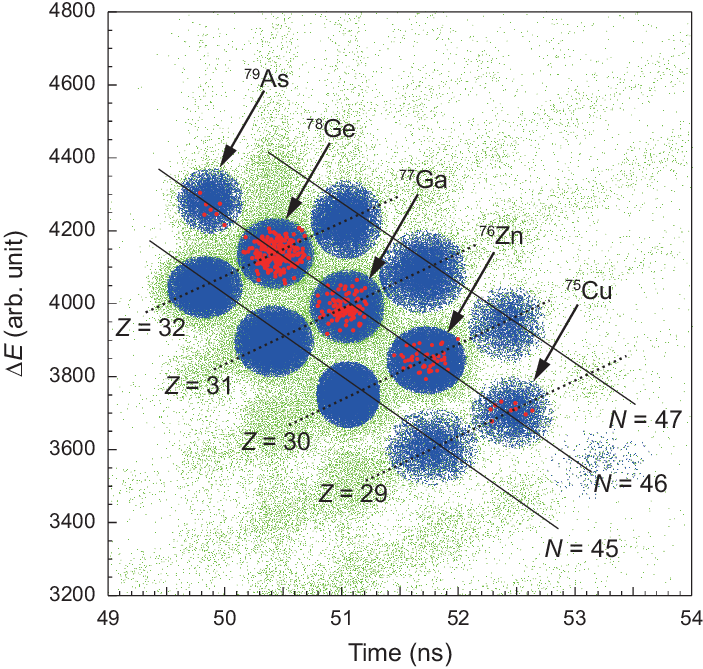}
\caption{Particle identification plot. 
The vertical and horizontal axes represent the energy deposit in the ionization chamber and the TOF between F2 and F3, respectively.
The green, blue, and red dots indicate all ions causing valid trigger events, 
namely the particles identified and selected at F3, and the ions injected, stored, extracted from the ring and selected at ELC, respectively.} 
\label{fig:pi}
\end{figure}

The energy deposit $\Delta E$ in the ionization chamber at F3 and the TOF between F2 and F3 were measured for particle identification.
A typical particle-identification plot is presented in Fig.~\ref{fig:pi}.
All the nuclei which generate the trigger signal are indicated by the green dots. To select the valid events, a Gaussian fitting was applied to the $\Delta E$ and TOF spectra.
In this analysis, the selection gates of $\pm 2.6$ standard deviations were applied for both the $\Delta E$ and TOF spectra, as indicated by the blue symbols in Fig.~\ref{fig:pi}.

\subsection{Injection of RIs into the ring}
The fission fragments identified at F3 were transported to the ring via the long-injection beamline. 
Several detectors were placed along it for beam diagnostics.
The momenta of the nuclei of interest were measured at the dispersive focal plane F6 of the BigRIPS by using a position-sensitive parallel-plate avalanche counter (PPAC)~\cite{kumagai}.
The momentum dispersion at F6 ($D_{\mathrm{F6}}$) was measured in this experiment to be $72.5(6)$~mm/\%.

\begin{figure}[htb]
\centering
\includegraphics[clip, width=12.9cm]{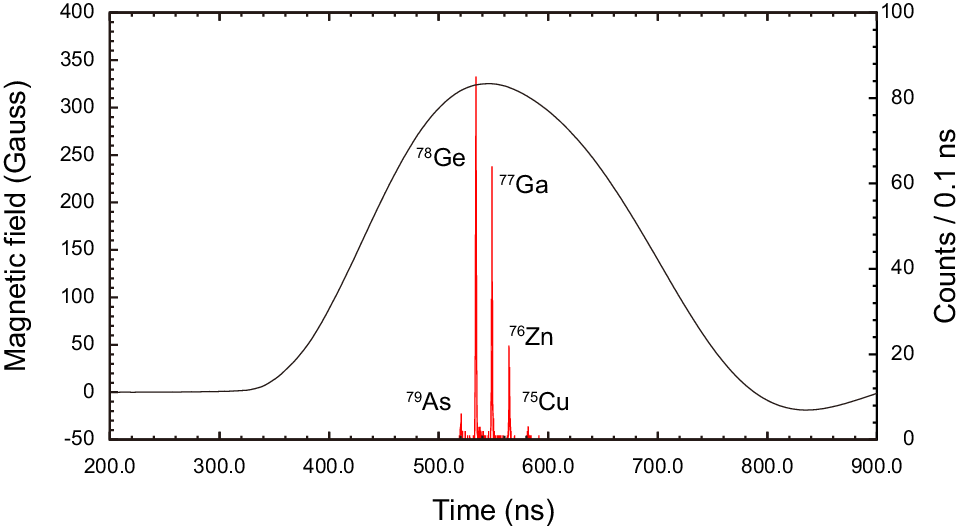}
\caption{Waveform of the kicker magnetic field and a histogram of the arrival times of particles of interest. Only the successfully injected particles are considered.
The zero of the time scale corresponds to the time when the trigger signal arrived at the kicker, that is the time at F3 with a constant offset of about 1~$\mu$s. 
The left and right vertical axes are the magnetic field and counting rates per 0.1~ns, respectively.}
\label{fig:injection}
\end{figure}

Afterwards, the fission fragments passed through the septum magnets and arrived at the kicker magnets. 
After the accomplishment of the beam tuning, the detectors placed in the long-injection beamline were removed,
except for those placed at F2, F3, F6, and S0. 
The fission fragments were then individually injected into the ring utilizing the fast-response kicker system~\cite{miura}.
The flight time of the ions from F3 to the kicker was approximately 1~$\mu$s.
The magnetic field of the kicker reached a flat top 950~ns after the trigger signal was generated at F3.
This includes the time delays needed for the signal propagation from F3 to the kicker, internal kicker electronic processing, and the response of the magnet itself.
To synchronize the particle arrival and the excitation of the magnetic field, an additional 50~ns delay was implemented.
A typical waveform of the kicker magnetic field and a histogram of arrival times of the ions of interest at R-MD4 are shown in Fig.~\ref{fig:injection}.
As a result, the entire range of the ions of interest could be injected in a single optimized timing setting. 

\subsection{Confirmation of circulation of RIs}
\label{subsec:cir}
\begin{figure}[htb]
\centering
\includegraphics[clip, width=12.9cm]{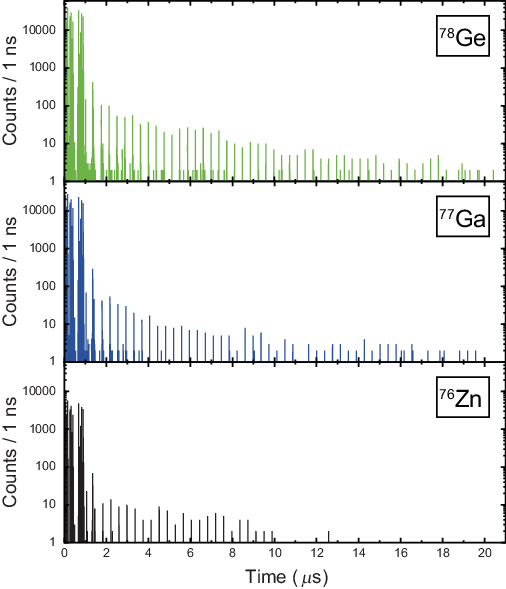}
\caption{Time signals of $^{78}$Ge, $^{77}$Ga, and $^{76}$Zn taken from Ref.~\cite{nagae}.
Noise at 0--1~$\mu$s is due to the kicker magnets. 
}
\label{fig:storage}
\end{figure}

Since only a single ion is injected at a time, standard beam diagnostics is insufficiently sensitive. 
The first challenge is to control whether an injected particle is revolving in the ring. 
The circulation of the stored particles was confirmed using an in-ring TOF detector, termed E-MCP, equipped with a carbon foil [E-MCP(C)]~\cite{nagae} placed at MCPC; see Fig.~\ref{fig:ring}.
The obtained time spectra of $^{78}$Ge, $^{77}$Ga, and $^{76}$Zn are shown in Fig.~\ref{fig:storage}.
The observed periodic time signals indicate that particle circulation was successfully achieved.
Stored $^{78}$Ge ions circulated for about 60 turns.

The corresponding revolution times were obtained through the analysis described in detail in Ref.~\cite{nagae}. 
The results are presented in Table~\ref{tab:revolution}.
The amounts of $^{79}$As and $^{75}$Cu were insufficient for deducing their revolution times directly.
Nevertheless, the revolution times for $^{79}$As and $^{75}$Cu can be deduced using Eq.~(\ref{eq:betaT}) and $^{78}$Ge as reference.
In this analysis, the $\beta$ values of $^{79}$As, $^{75}$Cu, and $^{78}$Ge can be calculated by taking the magnetic rigidity of $B \rho = 4.8513$~Tm as measured by an NMR probe located in the ring.
The calculated revolution times for $^{79}$As and $^{75}$Cu are also presented in Table~\ref{tab:revolution}.
After the circulation of the ions in the ring was confirmed, the destructive E-MCP(C) was removed from the ring aperture.

\begin{table}[htb]
\begin{center}
\caption{Measured and calculated revolution times of stored ions. See text for more details.}
  \begin{tabular}{ccc} \hline
    RI & Measured revolution time (ns) & Calculated revolution time (ns)\\  \hline
    $^{79}$As &            & 368.404\\
    $^{78}$Ge & 373.121(2) &  \\
    $^{77}$Ga & 378.192(3) &  \\
    $^{76}$Zn & 383.615(5) &  \\
    $^{75}$Cu &            & 389.448 \\ \hline 
  \end{tabular}
\label{tab:revolution}
\end{center}
\end{table}

\subsection{Fine-tuning of the isochronous ion optics}
The quality of the isochronous ion optics is decisive for mass determination in isochronous mass spectrometry.
The optics is extremely sensitive to the applied magnetic fields and their quality.
The currents of the trim coils incorporated into the dipole magnets of the ring were tuned to obtain the required magnetic-field distribution.
The magnetic field setting was optimized for the reference particle, $^{78}$Ge, by monitoring the correlation between the TOF from S0 to ELC (TOF$_\mathrm{S0ELC}$) and the particle momentum, which was measured at F6 prior to the injection into the ring. 
If isochronous ion optics is perfectly optimized for the reference particle $^{78}$Ge, the TOF$_\mathrm{S0ELC}$($^{78}$Ge) will be constant and independent of momentum. 
Thus, the quality of the isochronous ion optics is quantified by the momentum dependence on TOF$_\mathrm{S0ELC}$($^{78}$Ge).
In this experiment, second-order ion-optical tuning was adopted.

\begin{figure}[htb]
\centering
\includegraphics[clip, width=12.9cm]{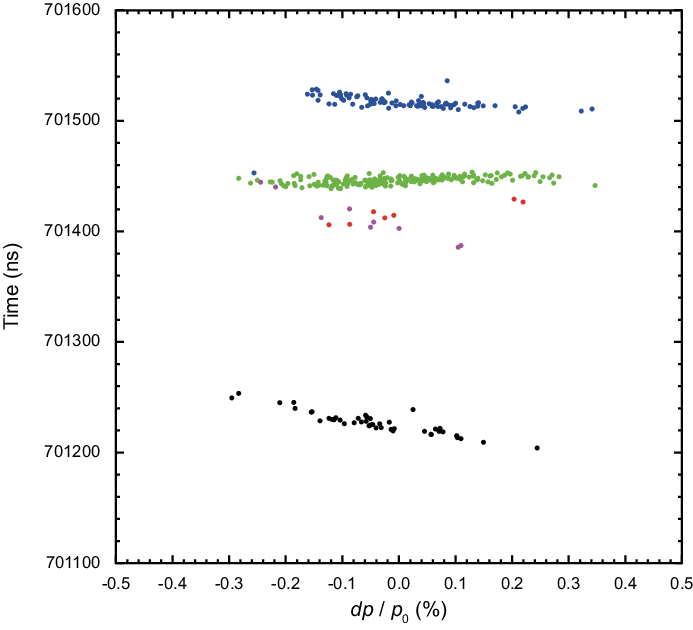}
\caption{Momentum dependence of the 
total flight time TOF$_\mathrm{S0ELC}$. The red, green, blue, black, and purple circles indicate $^{79}$As, $^{78}$Ge, $^{77}$Ga, $^{76}$Zn, and $^{75}$Cu, respectively.}
\label{fig:tofring}
\end{figure}

Green symbols in Fig.~\ref{fig:tofring} demonstrate the momentum dependence of the TOF$_\mathrm{S0ELC}$ of $^{78}$Ge reference particles.
In this figure, the abscissa axis is the momentum ($p$) deviation from the momentum at a weighted average position at F6 ($p_{0}$), $dp/p_{0} = (p - p_{0})/p_{0}$.
The momentum deviation was calculated using the position information of the events at F6 ($P_{\mathrm{F6}}$), the weighted average of $P_{\mathrm{F6}}$ ($P_{\mathrm{F6avg}}$) described in Sec.~\ref{subsec:beta}, and the momentum dispersion at F6 ($D_{\mathrm{F6}}$) as $dp/p_{0} = (P_{\mathrm{F6}} - P_{\mathrm{F6avg}})/D_{\mathrm{F6}}$.
The E-MCP(C) at S0 and the plastic scintillator at ELC were used to measure TOF$_\mathrm{S0ELC}$. 
Owing to the second-order tuning of the magnetic fields, 
a small third-order dependence appeared in the correlation between TOF$_\mathrm{S0ELC}$($^{78}$Ge) and the momentum.
{A precision of TOF$_\mathrm{S0ELC}$($^{78}$Ge) was achieved to be $4.7$~ppm (standard deviation) for a momentum spread of $\pm 0.3$\%.}

\subsection{Storage and extraction of RIs and TOF measurement}
After a defined storage time, the particles were extracted from the ring using the same kicker magnets and the same magnetic field distribution as the one used for the injection.
Two storage-time intervals were set, namely $700.6$~$\mu$s to extract $^{79}$As, $^{78}$Ge, $^{77}$Ga, and $^{75}$Cu, 
and $700.4$~$\mu$s for $^{76}$Zn.

Total flight time TOF$_\mathrm{S0ELC}$ spectra are shown in Fig.~\ref{fig:tofring}. 
In contrast to the TOF$_\mathrm{S0ELC}$ of  $^{78}$Ge, momentum dependences are clearly observed in TOF$_\mathrm{S0ELC}$ for $^{79}$As, $^{77}$Ga, $^{76}$Zn, and $^{75}$Cu.
The TOF$_\mathrm{S0ELC}$ of approximately 0.7~ms was used instead of measuring revolution times in the ring turn by turn.
The latter would be advantageous, but the corresponding detector needs to be non-destructive, which remains a challenge.
The previously discussed E-MCP(C) counter introduces energy losses in the carbon foil, which limits the particle storage to only a few tens of turns, whereas about 2000 revolutions are needed to achieve the anticipated mass precision.

The number of turns for each particle was determined using the revolution time and the time interval between injection and extraction.
The obtained turn numbers $N$ are listed in Table~\ref{tab:turn}.
In the mass determination, the listed turn numbers were incremented by 2 in consideration that the sum of the flight path lengths from S0 to the kicker and from the kicker to ELC is approximately twice the ring circumference.

A long-range time-to-digital converter (Acqiris TC842~\cite{acqiris}) was used to measure the total TOF$_\mathrm{S0ELC}$ defined by the start signal from the E-MCP(C) at S0 and the stop signal from the plastic scintillator at ELC.
The signals from the E-MCP(C), plastic scintillator, and the trigger detector were processed through a constant fraction discrimination (Ortec935~\cite{ortec935}) before being fed into the Acqiris TC842. 

\begin{table}[htb]
\begin{center}
\caption{Numbers of turns, $N$, of stored $^{79}$As, $^{78}$Ge, $^{77}$Ga, $^{76}$Zn, and $^{75}$Cu. See text for more details.}
  \begin{tabular}{cc} \hline
    RI  & Number of turns $N$ \\ \hline
    $^{79}$As & 1902 \\
    $^{78}$Ge & 1878 \\
    $^{77}$Ga & 1853 \\
    $^{76}$Zn & 1826 \\
    $^{75}$Cu & 1799 \\ \hline
  \end{tabular}
\label{tab:turn}
\end{center}
\end{table}

\section{Mass determination}
To determine $m_{1}/q_{1}$ using Eqs.~(\ref{eq:m/q-beta}) and (\ref{eq:m/q-brho}), 
the revolution times $T_{0, 1}$ and the velocity $\beta_{1}$ or  magnetic rigidity $B\rho$ need to be known.
The revolution times $T_{0, 1}$ were calculated using the corresponding TOF$_\mathrm{S0ELC}$ and $N$.
Since $T_{1}$ can be corrected by using either $\beta_{1}$ or $B\rho$, there are two analytical methods for determining $m_{1}/q_{1}$.

\subsection{Extracted events selection}
\begin{figure}[tb]
\centering
\includegraphics[clip, width=12.9cm]{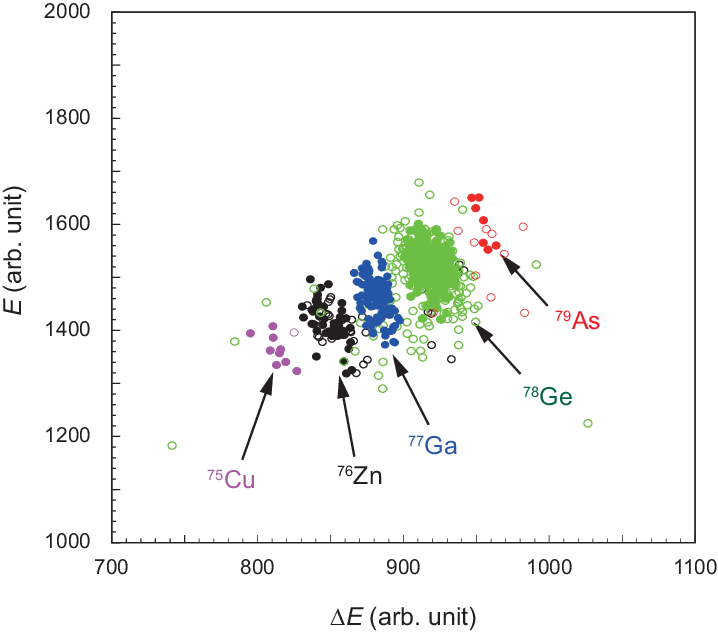}
\caption{Particle identification plot at ELC. 
The open and filled circles indicate the events selected at F3 and the events selected at ELC, respectively.}
\label{fig:ELC-PI}
\end{figure}

As described in Sec.~\ref{subsec:PI}, the fission fragments were identified at F3 prior to injection. Furthermore, the fragments were also identified at ELC as successfully extracted events.
The extracted events were defined as those that had proper TOF stop data from the plastic scintillator at ELC.
In addition, events were selected from a particle identification at ELC using the total energy $E$ and $\Delta E$ information obtained by the NaI(Tl) counter and ionization chamber, respectively, as shown in Fig~\ref{fig:ELC-PI}.
The selection gates of $\pm 3.0$ and $\pm 1.5$ standard deviations were applied to both the $E$ and $\Delta E$ spectra, respectively.
The selected valid extraction events obtained in a 3-hour-accumulation period are plotted and indicated by the red dots in Fig.~\ref{fig:pi}.
The number of selected extraction events is listed in Table~\ref{tab:event}.

\begin{table}[htb]
\begin{center}
\caption{Numbers of extracted events of $^{79}$As, $^{78}$Ge, $^{77}$Ga, $^{76}$Zn, and $^{75}$Cu. See text for more details.}
  \begin{tabular}{cS} \hline
    {RI}  & {Number of events} \\ \hline
    {$^{79}$As} &   7 \\
    {$^{78}$Ge} & 218 \\
    {$^{77}$Ga} &  94 \\
    {$^{76}$Zn} &  45 \\
    {$^{75}$Cu} &   9 \\ \hline
  \end{tabular}
\label{tab:event}
\end{center}
\end{table}

\subsection{Velocity determination}
\label{subsec:beta}
\begin{figure}[htb]
\centering
\includegraphics[clip, width=12.9cm]{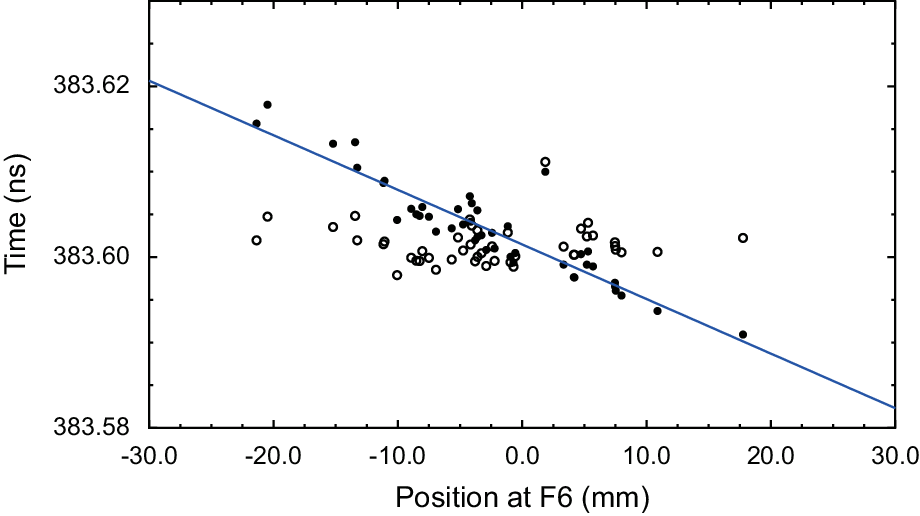}
\caption{Correlation between $T_{1}$ ($T_{1\mathrm{-cor}}$) and the position at F6 for $^{76}$Zn.
The black filled circles indicate $T_{1}$ and the blue line represents a straight-line fit result.
The black open circles indicate $T_{1\mathrm{-cor}}$.}
\label{fig:F6p-T1}
\end{figure}

Absolute values of velocity $\beta$ are crucial for correcting $T_{1}$ via Eq.~(\ref{eq:m/q-beta}). 
Each $\beta$ value was calculated employing TOF from F3 to S0, TOF$_{\mathrm{F3S0}}$, as:
\begin{equation}
\label{eq:beta-VRA}
\beta  = \frac{L_\mathrm{F3S0}}{c \left( \mathrm{TOF_{F3S0}} + O \right)},
\end{equation}
where $L_\mathrm{F3S0}$ is the flight path length between F3 and S0, and
$O$ is the offset of the measured TOF$_\mathrm{F3S0}$.
To obtain $L_\mathrm{F3S0}$ and $O$,
we assume that the velocities between F3 and S0, $v_\mathrm{F3S0}$, and between S0 and ELC, $v_\mathrm{S0ELC}$, are the same.
Thus, the following equations hold:
\begin{equation}
\label{eq:velocity-VRA}
\begin{aligned}
\frac{L_\mathrm{F3S0}}{\mathrm{TOF_{F3S0}} + O} & = \frac{L_\mathrm{ring}}{T_{0,1}},\\
\mathrm{TOF_{F3S0}} & = \frac{L_\mathrm{F3S0}}{L_\mathrm{ring}}T_{0,1} - O,
\end{aligned}
\end{equation}
where $L_\mathrm{ring}= 60.335(2)$~m is the flight length in the ring, see Sec.~\ref{subsec:Brho}.
To obtain the precise velocities,
we used momentum-corrected TOF$_\mathrm{F3S0-cor}$ and $T_{0,1 \mathrm{-cor}}$ 
instead of TOF$_\mathrm{F3S0}$ and $T_{0, 1}$ in Eq.~(\ref{eq:velocity-VRA}) using the following equations:
\begin{equation}
\label{eq:T-cor}
\begin{aligned}
\mathrm{TOF_{F3S0-cor}}  & = \mathrm{TOF_{F3S0}}  -a \left( P_\mathrm{F6} - P_\mathrm{F6avg} \right),\\
T_{0,1 \mathrm{-cor}} & = T_{0,1} -a' \left( P_\mathrm{F6} - P_\mathrm{F6avg} \right),
\end{aligned}
\end{equation}
where $a$ ($a'$) denotes the slope of the TOF$_\mathrm{F3S0}$--position(momentum) ($T_{0,1}$--position) correlation.
The slopes of the correlations were obtained by a fitting analysis using a first-order polynomial.
An example of the correlation plot of $T_{1}$ for $^{76}$Zn is shown by the black filled circles in Fig.~\ref{fig:F6p-T1}.
The momentum dependence of $T_{1}$ is clearly seen and can be corrected accordingly.
The corrected $T_{1}$, $T_{1\mathrm{-cor}}$, is indicated by the black open circles in Fig.~\ref{fig:F6p-T1}. 
The momentum dependence was sufficiently reduced through this procedure. 
The same analysis was performed for $^{79}$As, $^{78}$Ge, $^{77}$Ga, and $^{76}$Zn.

\begin{figure}[htb]
\centering
\includegraphics[clip, width=8.6cm]{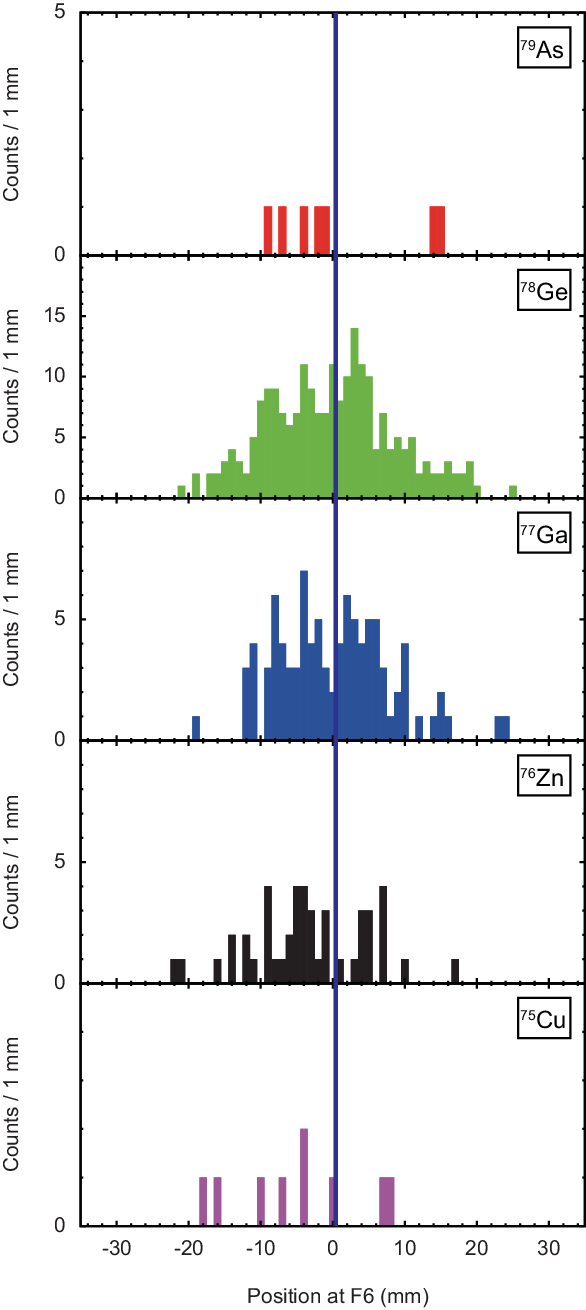}
\caption{Position distribution of each isotope at F6. The blue line indicates the weighted average position ($P_\mathrm{F6avg}$). See text for more details.}
\label{fig:f6x}
\end{figure}

Figure~\ref{fig:f6x} depicts the position distributions for each stored nuclear species at the dispersive focal plane F6.
The weighted average position, $P_\mathrm{F6avg} = 0.04(44)$~mm, is indicated by the blue line in Fig.~\ref{fig:f6x}, which was obtained
by considering the weighted average of the individual weighted average positions of $^{79}$As, $^{78}$Ge, $^{77}$Ga, and $^{76}$Zn.

\begin{figure}[htb]
\centering
\includegraphics[clip, width=12.9cm]{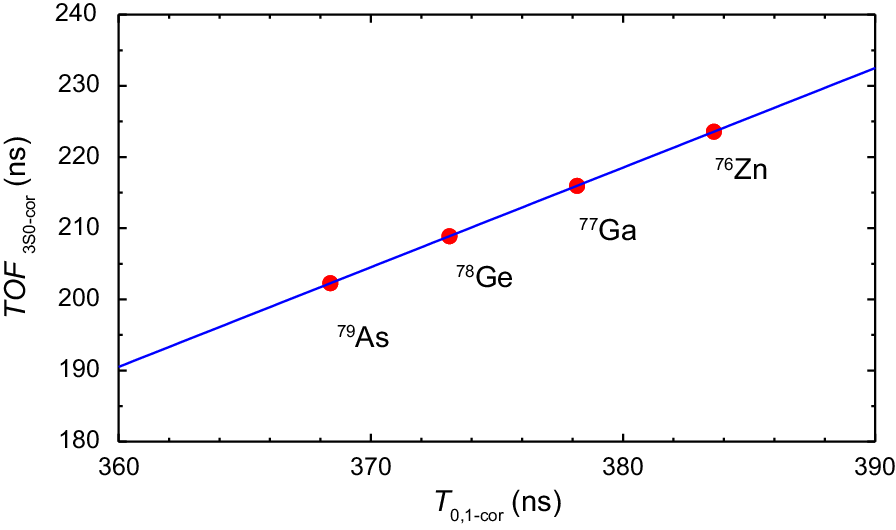}
\caption{Relationship between TOF$_\mathrm{F3S0-cor}$ and $T_{0,1 \mathrm{-cor}}$. 
The blue line is a linear fit result.
All error bars are smaller than the symbol size.}
\label{fig:beta}
\end{figure}

$L_\mathrm{F3S0}$ and $O$ were determined 
through a linear fit according to Eq.~(\ref{eq:velocity-VRA}) of the correlation plot between TOF$_\mathrm{F3S0-cor}$ and $T_{0,1 \mathrm{-cor}}$, as shown in Fig.~\ref{fig:beta}.
The uncertainties of TOF$_\mathrm{F3S0-cor}$ were about 0.3 to 1.6~ps, while the fit residua [TOF$_\mathrm{F3S0-cor}(\mathrm{exp}) - $TOF$_\mathrm{F3S0-cor}(\mathrm{fit})$] were within $\pm 10$~ps.
The resultant $L_\mathrm{F3S0}$ and $O$ values are $L_\mathrm{F3S0} = 84.465(5)$~m and $O = 313.47(2)$~ns.
The $\beta$ values of $^{75}$Cu were calculated event-by-event using Eq.~(\ref{eq:beta-VRA}) adopting the measured TOF$_\mathrm{F3S0}$, flight length $L_\mathrm{F3S0}$, and offset $O$.

\begin{figure}[htb]
\centering
\includegraphics[clip, width=12.9cm]{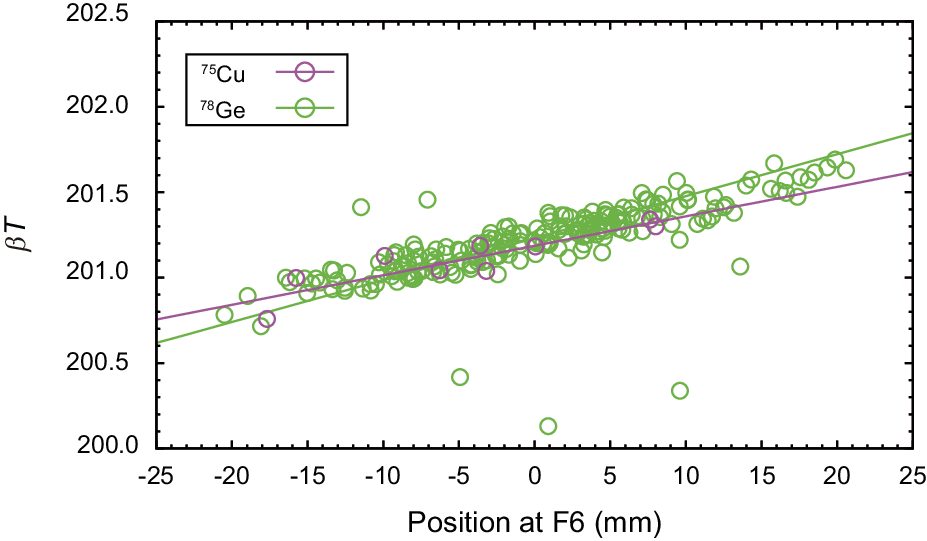}
\caption{$\beta T$ as a function of position at F6 for $^{75}$Cu and $^{78}$Ge. The solid lines are fit results. See text for more details.
}
\label{fig:betaT}
\end{figure}

Based on the principle of mass measurement, $\beta_{1} T_{1}$ of $^{75}$Cu should coincide with $\beta_{0} T_{0}$.
However, the experimental $\beta_{1} T_{1}$ of $^{75}$Cu exhibited a small difference.
Possible reasons for this disagreement may be the momentum-dependent flight path lengths and energy losses in the PPAC at F6, which could not be accurately considered. 
These effects are sources of systematic uncertainties.
To account for them, we introduced a correction factor for $\beta_{1}$ of $^{75}$Cu to match $\beta_{0} T_{0}$.
The correction factor is defined as the ratio $R = \beta_{0} T_{0} / \beta_{1} T_{1} (^{75} \mathrm{Cu})$ at $P_{\mathrm{F6avg}}$.
To obtain the ratio $R$, the experimental $\beta_{0} T_{0}$ and $\beta_{1} T_{1} (^{75} \mathrm{Cu})$ as a function of the F6 position were fitted using the first-order polynomial,
as shown in Fig.~\ref{fig:betaT}.
The fit results of the $\beta T$ values at the F6 weighted average position were used to calculate the ratio $R$ giving $R = 1.00023(20)$ for $^{75}$Cu.
The final $\beta$ taken for the mass determination was obtained via $\beta_{1 \mathrm{-cor}} = R \beta_{1}$.

\clearpage

\subsection{Magnetic rigidity determination}
\label{subsec:Brho}
The absolute magnetic rigidities of each particle are crucial for correcting $T_{1}$ via Eq.~(\ref{eq:m/q-brho}).
The magnetic rigidities were not directly measured for each event and were thus calculated using the following equation:
\begin{equation}
\label{eq:brho-ring}
\begin{aligned}
B \rho = B \rho_\mathrm{ring}  \left\{ 1 + \frac{\left( P_\mathrm{F6} - P_\mathrm{F6avg} \right) }{D_\mathrm{F6}}   \right\},
\end{aligned}
\end{equation}
where $B\rho_\mathrm{ring}$ is the mean value of the magnetic rigidity of the ring, $P_\mathrm{F6}$ is the F6 position, $P_\mathrm{F6avg}$ is the weighted average position at F6, and $D_\mathrm{F6} = 72.5(6)$~mm/$\%$ is the dispersion measured at F6.

Based on the definition of magnetic rigidity, the relationship between $m/q$ (u/$q$) and $B\rho$ is:
\begin{equation}
\label{eq:mq-brho}
\begin{aligned}
\frac{m}{q} = \frac{B\rho} {\gamma v} = \frac{B\rho \sqrt{1-(v/c)^2}}{v},\\
\end{aligned}
\end{equation}
where u  and $v$ denote the unified atomic mass unit, and velocity, respectively.
The velocity, respectively can be expressed as $l / t$ with a flight length $l$ and time-of-flight $t$. 
Hence, the relationship between $t$ and $B\rho$ is obtained:
\begin{equation}
\label{eq:T'2-Brho}
\begin{aligned}
t^2 = \left( \frac{l}{B\rho} \right)^2  \left( \frac{m}{q}\right)^2  + \left( \frac{l}{c} \right)^2.
\end{aligned}
\end{equation}

\begin{figure}[htb]
\centering
\includegraphics[clip, width=12.9cm]{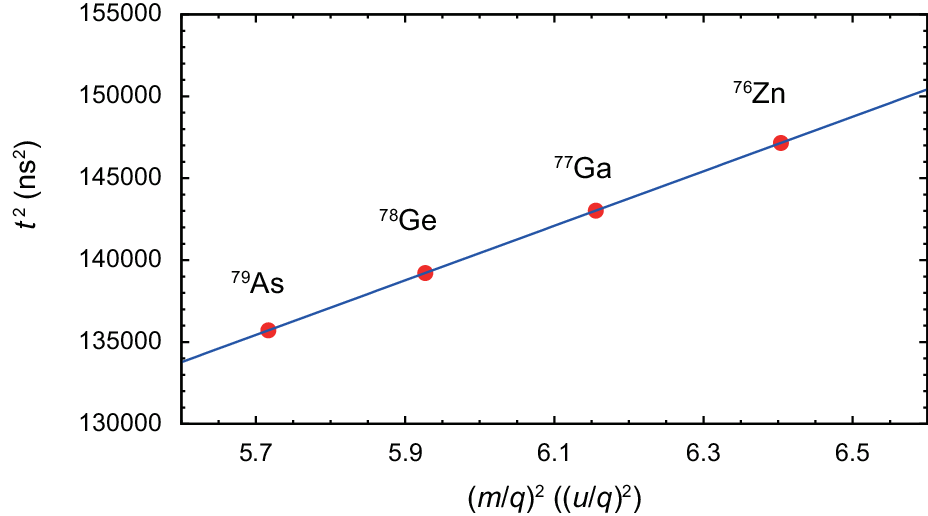}
\caption{Relationship between $t^2$ and $(m/q)^2$ according to Eq.~(\ref{eq:T'2-Brho}). 
The blue line indicates the result of the fitting analysis.
The error bars of $t^{2}$ are smaller than the symbol size.
}
\label{fig:brho}
\end{figure}

To evaluate $B\rho_\mathrm{ring}$, 
the momentum-corrected revolution times $T_{0,1 \mathrm{-cor}}$ and the mean ring circumference $L_\mathrm{ring}$ are taken as $t$  and $l$, respectively, in Eq.~(\ref{eq:T'2-Brho}).
Figure~\ref{fig:brho} presents $t^2$ as a function of the squared mass-to-charge ratio $(m/q)^2$ from AME2020~\cite{ame}.
A linear fitting analysis using Eq.~(\ref{eq:T'2-Brho}) provided $B\rho = B\rho_\mathrm{ring}$ and $l = L_\mathrm{ring}$ as free parameters.
Consequently, $B\rho_\mathrm{ring}$ and $L_\mathrm{ring}$ were obtained to be $B\rho_\mathrm{ring} = 4.8457(2)$~Tm and  $L_\mathrm{ring} = 60.335(2)$~m.
Whereas the uncertainties of the $t^2$ data points are about 0.1 to 0.3~ns$^2$, the corresponding uncertainties of the fit residua [$t^2(\mathrm{exp}) - t^2(\mathrm{fit})$] are as small as $\pm 0.1$~ns$^2$.
The obtained magnetic field in the analysis was different from that measured by the NMR probe in the ring described in Sec.~\ref{subsec:cir}.
This difference could be attributed to the fact that the NMR probe in the ring was not installed at a position corresponding to the momentum at $P_{\rm F6avg}$.
The magnetic rigidities of each particle were determined through Eq.~(\ref{eq:brho-ring}).

\subsection{Revolution times in the Rare-RI Ring}
\begin{figure}[htb]
\centering
\includegraphics[clip, width=12.9cm]{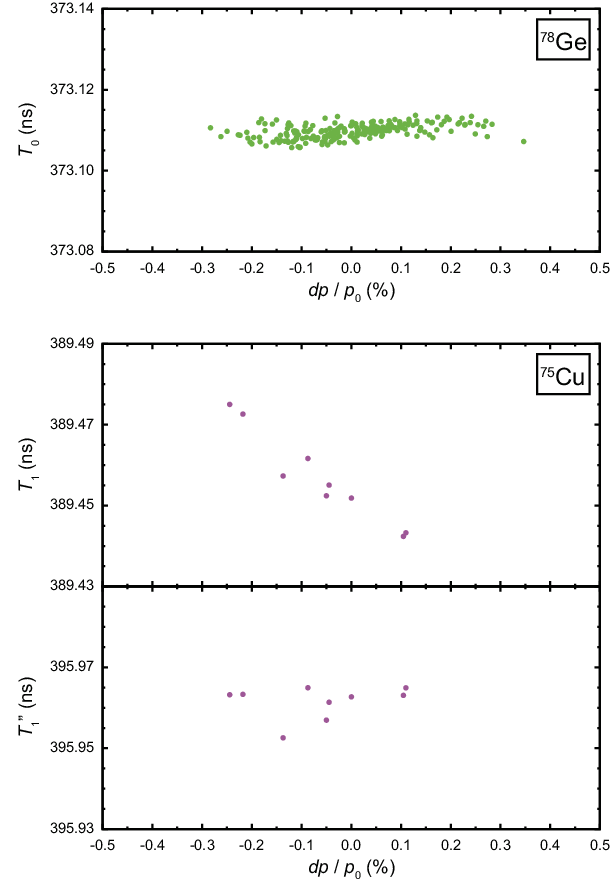}
\caption{Top and middle: Relationship between $T_{0,1}$ as a function of the momentum. Bottom: $B\rho$ corrected $T_{1}''$ of $^{75}$Cu.}
\label{fig:F6pTOFring}
\end{figure}

The revolution times $T_{0, 1} = \mathrm{TOF_{S0ELC}}/ (N+2)$
obtained for $^{78}$Ge and $^{75}$Cu are shown in Fig.~\ref{fig:F6pTOFring} 
as a function of the momentum measured at F6. 
Because isochronism is fulfilled for the reference $^{78}$Ge,
the momentum dependence of $T_{0}$ of $^{78}$Ge is not observed; however, that of $^{75}$Cu is apparent and can be corrected using $\beta$ or $B\rho$.
Figure~\ref{fig:F6pTOFring} (bottom) presents $T_{1}''$, which is corrected by $B\rho$.
The mean $T_{0}$ value was $373.1098(1)$~ns, and the mean $T_{1}''$ value was $395.961(1)$~ns.
Therefore, the uncertainty of $T_{0}$ is on the order of $10^{-7}$ and that of $T_{1}''$ is on the order of $10^{-6}$. 
The uncertainty of $T_{0}$ mainly stems from the magnetic field instabilities and time resolutions of the timing detectors at S0 and ELC. 

\clearpage

\subsection{Masses}
\begin{figure}[htb]
\centering
\includegraphics[clip, width=12.9cm]{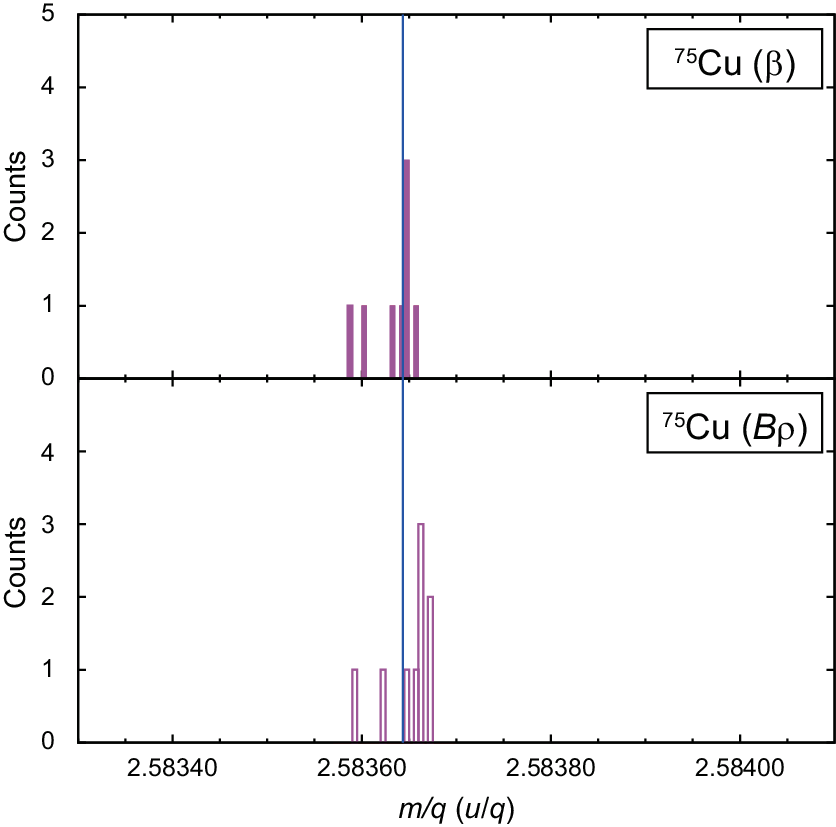}
\caption{$m_{1}/q_{1}$ spectra of $^{75}$Cu obtained adopting the $\beta$ and $B\rho$ corrections. The blue line indicates the literature value.}
\label{fig:moq-spe}
\end{figure}

\begin{figure}[htb]
\centering
\includegraphics[clip, width=12.9cm]{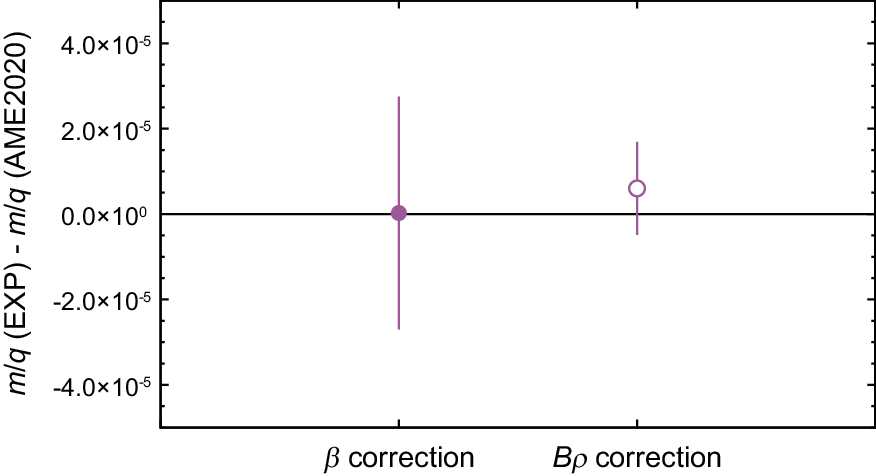}
\caption{Difference of the experimental $^{75}$Cu $m/q$ values for two correction methods and the literature value.}
\label{fig:moq-dif}
\end{figure}

The $m_{1}/q_{1}$ values were evaluated for each event by adopting the corresponding $T_{1}$, $\beta$, 
and $B\rho$ in Eqs.~(\ref{eq:m/q-beta}) and (\ref{eq:m/q-brho}).
In the analyses, a constant $T_{0}$ value of $373.1098$~ns was used.
Because the mass of $^{75}$Cu is well known, comparing the obtained experimental values to the literature values from AME2020~\cite{ame} allows the verification of the present mass determination procedure. 
Figure~\ref{fig:moq-spe} presents the $m_{1}/q_{1}$ spectra of $^{75}$Cu obtained using the $\beta$ and $B\rho$ corrections.
The $m_{1}/q_{1}$ value  $m_{1}/q_{1} (\beta) = 2.583644(27)$~u/$q$ was obtained by employing the $\beta$ correction and $m_{1}/q_{1} (B\rho) = 2.583650(11)$~u/$q$ by employing the $B\rho$ correction.
Figure~\ref{fig:moq-dif} presents a comparison of the experimental $m_{1}/q_{1}$ values with the literature $m/q$ value [$m_{1}/q_{1} (\mathrm{AME2020}) = 2.58364357(3)$~u/$q$].
The differences between the experimental and literature values is less than $1 \times 10^{-5}$~u/$q$.
The mass excess ($ME$) values obtained in these measurements are $ME (\beta) = -$54460(740)~keV and $ME (B\rho) = -$54310(290)~keV, while the literature mass excess value is $ME (\mathrm{AME2020}) = -$54470.2(7)~keV~\cite{ame}.

The total uncertainty for the $\beta$ correction analysis was deduced via
\begin{equation}
\label{eq:error-beta}
\delta \left( \frac{m_{1}}{q_{1}} \right)_{\beta}  = \sqrt{
\left(\delta \frac{m_{1}}{q_{1}} \right)^2_{\mathrm{sta}} + \left(\delta \frac{m_{1}}{q_{1}} \right)^2_{T_{0}} 
 + \left(\delta \frac{m_{1}}{q_{1}} \right)^2_{\beta_{1}} + \left(\delta \frac{m_{1}}{q_{1}} \right)^2_{R}
},
\end{equation}
where $\delta (m_1/q_1)_{\mathrm{sta}}$ is the statistical uncertainty of $m_{1}/q_{1}$, and $\delta (m_1/q_1)_{T_{0}}$, $\delta (m_1/q_1)_{\beta_{1}}$, and $\delta (m_1/q_1)_{R}$ are propagated statistical uncertainties from $T_{0}$, $\beta_{1}$ , and $R$ corrections, respectively.
For the $B \rho$ correction analysis, the total uncertainty is given as
\begin{equation}
\label{eq:error-Brho}
\begin{aligned}
\delta \left( \frac{m_{1}}{q_{1}} \right)_{B\rho} & = \sqrt{
\left(\delta \frac{m_{1}}{q_{1}} \right)^2_{\mathrm{sta}} + \left(\delta \frac{m_{1}}{q_{1}} \right)^2_{T_{0}} + \left(\delta \frac{m_{1}}{q_{1}} \right)^2_{B\rho_{\mathrm{ring}}} 
+ \left(\delta \frac{m_{1}}{q_{1}} \right)^2_{D_{\mathrm{F6}}} + \left(\delta \frac{m_{1}}{q_{1}} \right)^2_{P_{\mathrm{F6avg}}}
},
\end{aligned}
\end{equation}
where $\delta (m_1/q_1)_{B\rho_{\mathrm{ring}}}$, $\delta (m_1/q_1)_{D_{\mathrm{F6}}}$, and $\delta (m_1/q_1)_{P_{\mathrm{F6avg}}}$ are the uncertainties due to the error propagation from $B\rho_{\mathrm{ring}}$, $D_{\mathrm{F6}}$, and $P_{\mathrm{F6avg}}$, respectively.
Because the relative uncertainty of $\delta (m_{0}/q_{0}) / (m_{0}/q_{0})$ is negligible ($\approx 3 \times 10^{-8}$), 
we neglected this contribution in the analyses.

The statistical uncertainties of $m_{1}/q_{1}$ values were obtained using the standard deviation, $\sigma$, of the $m_{1}/q_{1}$ distribution and the number of events, $n$,
using the equation $\delta (m_{1} / q_{1})_{\mathrm{sta}} = \sigma / \sqrt{n}$.
These uncertainties were deduced to be $\delta (m_{1} / q_{1})_{\mathrm{sta}} (\beta) = 1.3 \times 10^{-5}$~u/$q$ and $\delta (m_{1} / q_{1})_{\mathrm{sta}} (B\rho) = 8.9 \times 10^{-6}$~u/$q$ for the $\beta$ and $B\rho$ corrections, respectively.
The time fluctuations of the ring magnetic field change the time-of-flight in the ring and broaden the peak width of the $m/q$ spectrum. As a result, it affects the statistical uncertainty of $m_{1}/q_{1}$.

\begin{figure}[htb]
\centering
\includegraphics[clip, width=12.9cm]{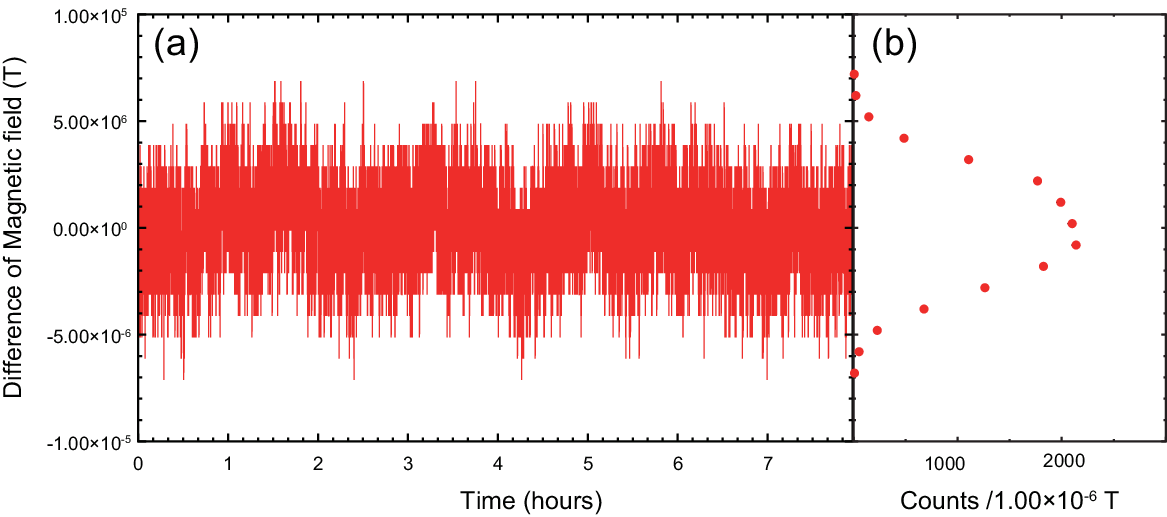}
\caption{(a) Time variation of the magnetic field. The differences from the average magnetic field are plotted.
(b) The projection of Fig.~\ref{fig:NMR} (a).
}
\label{fig:NMR}
\end{figure}

Figure~\ref{fig:NMR} (a) presents the typical time fluctuations of the magnetic field of a dipole magnet during the measurement period.
The magnitude of the fluctuation shown in Fig.~\ref{fig:NMR} (b) was estimated to be $2.3 \times 10^{-6}$~T (standard deviation), while the average magnetic field was 1.2035462~T.
A momentum dependence of the $T_{1}'$ and $T_{1}''$ values also affects the statistical uncertainty of $m_{1}/q_{1}$.
As shown in Fig.~\ref{fig:F6pTOFring}(bottom), the momentum dependence of the $T_{1}''$ value remained observable.
This dependence arises from the inaccurately known $\beta$ or $B\rho$.

The uncertainty of $T_{0}$ depends on the quality of isochronous optics and timing resolution of the detectors. 
The main contribution to the uncertainty of $T_{0}$ stems from the optics imperfections,
because the time resolution of the detectors was on the order of several tens of ps. 
The uncertainties were deduced to be $\delta (m_{1}/q_{1})_{T_{0}} = 1.2 \times 10^{-6}$~u/$q$ and $1.1 \times 10^{-6}$~u/$q$ for the $\beta$  and $B\rho$ corrections, respectively. 
The uncertainties of $\beta$, $R$, $B\rho_{\mathrm{ring}}$, $D_{\mathrm{F6}}$, and $P_{\mathrm{F6avg}}$ depend on the statistics and fitting procedure.
The contributions of each uncertainty and the total uncertainties are listed in Table~\ref{tab:error-slm}.

The total obtained uncertainties were $\delta (m_{1}/q_{1})_{\beta} = 2.7 \times 10^{-5}$~u/$q$ and $\delta (m_{1}/q_{1})_{B\rho} = 1.1 \times 10^{-5}$~u/$q$ for the $\beta$ and $B\rho$ correction approaches, respectively.
These values correspond to $\delta (m_{1})_{\beta} = 740$~keV and $\delta (m_{1})_{B\rho} = 290$~keV.
These uncertainties are about 10 times larger than the uncertainties of the upgraded isochronous mass measurements in CSRe~\cite{Meng-2022, Zhang-2023, Zhou-2023}, but comparable to the uncertainties of the original isochronous mass measurements in ESR~\cite{knobel-esr-ims, Knobel-2016b, Stadlmann-2004}.

Significant contributions to the total uncertainties were due to the statistical uncertainty of $m_{1}/q_{1}$ and uncertainty arising from the correction factor $R$ in the $\beta$ correction procedure.
For the $B \rho$ correction routine, the statistical uncertainty $m_{1}/q_{1}$ was the main contributor to the total uncertainty.
These results suggest that the improvement in the instabilities of the magnetic fields and the determination of accurate $\beta$ and $B\rho$ values are essential.
In this study, simple methods to obtain $\beta$ and $B\rho$ were adopted.
However, the analysis can be further improved to obtain more accurate $\beta$ and/or $B\rho$.

\begin{table}[htb]
\begin{center}
\caption{Uncertainties considered in the determination of the $m_{1}/q_{1}$ value for $^{75}$Cu. Heading $\beta$ and $B\rho$ stand for the employed correction method.}
\begin{tabular}{lcc}\hline
    Contribution                              & $\beta$~(u/$q$)               & $B\rho$~(u/$q$) \\ \hline
    $\delta (m_{1}/q_{1})_{\mathrm{sta}}$         & $1.3 \times 10^{-5}$           & $8.9 \times 10^{-6}$ \\
    $\delta (m_{1}/q_{1})_{T_{0}}$                & $1.2 \times 10^{-6}$          & $1.1 \times 10^{-6}$ \\
    $\delta (m_{1}/q_{1})_{\beta}$               & $6.7 \times 10^{-6}$           & -- \\
    $\delta (m_{1}/q_{1})_{R}$                  & $2.3 \times 10^{-5}$           & -- \\
    $\delta (m_{1}/q_{1})_{B\rho_\mathrm{ring}}$     & --                            & $3.3 \times 10^{-6}$ \\
    $\delta (m_{1}/q_{1})_{D_\mathrm{F6}}$          & --                            & $4.3 \times 10^{-7}$ \\
    $\delta (m_{1}/q_{1})_{P_\mathrm{F6avg}}$       & --                            & $5.1 \times 10^{-6}$\\ \hline
    $\delta (m_{1}/q_{1})$                      & $2.7 \times 10^{-5}$           & $1.1 \times 10^{-5}$ \\ \hline
  \end{tabular}
\label{tab:error-slm}
\end{center}
\end{table}

\clearpage

\section{Conclusion}
This study describes the first isochronous mass spectrometry implemented in the Rare-RI Ring,
which was conducted on nuclei with well-known masses and preceded the first mass measurements of new results reported in Ref.~\cite{Li-RIKEN}.
Individual-ion injection method, ion injection, storage, and extraction, together with TOF measurements in the ring applied to exotic nuclei have successfully been achieved.
Consequently, the known mass of $^{75}$Cu was re-measured under the strict conditions of short measurement durations.
The measurement time of approximately 0.7~ms is excellently suited for pinning down yet unmeasured masses of the most exotic nuclei.

We established two analytical methods for the mass determination employing either $\beta$ or $B\rho$ corrections.
The obtained mass accuracies and uncertainties were on the order of the $10^{-6}$~u/$q$ and $10^{-5}$~u/$q$, respectively for both, $\beta$ and $B\rho$, correction methods.
The major sources of the total uncertainty for the $\beta$ correction analysis were the statistical uncertainty of $m_{1}/q_{1}$ and the uncertainty of 
$R$; whereas for the $B\rho$ correction analysis, it was the statistical uncertainty of $m_{1}/q_{1}$.
The instabilities of the magnetic fields of the Rare-RI Ring and the momentum dependence of $m_{1}/q_{1}$ affected the statistical uncertainty of the $m_{1}/q_{1}$ value.
This study suggests further improvements of the stability of the magnetic fields of the Rare-RI Ring to facilitate
 more accurate determination of $\beta$ and $B\rho$ values.

Nonetheless, the obtained mass of $^{75}$Cu is in satisfactory agreement with the literature value.
The Rare-RI Ring facility has proved to be a powerful tool for mass measurements of rare RIs.
Following the first results reported in \cite{Li-RIKEN}, the Rare-RI Ring will be used for further mass measurements of short-lived nuclei, especially for nuclei involved in the $r$ process.

\section{Acknowledgments}
The authors thank the accelerator staff of RIKEN Ring Cyclotron for support during the experiment. 
This experiment on the Rare-RI Ring was performed at the RI beam factory operated 
by RIKEN Nishina Center and CNS, University of Tokyo under the Experimental Program MS-EXP16-10. 
This work was supported by the RIKEN Pioneering Project funding and JSPS KAKENHI Grants No. 26287036, No. 25105506, No. 15H00830, and No. 17H01123.
YAL received funding from the European Research Council (ERC) under the European Union's Horizon 2020 research and innovation programme (Grant Agreement No. 682841 ``ASTRUm'').

\bibliography{ms03-airXiv}

\end{document}